\documentclass{article}
\usepackage{spconf,amsmath,graphicx}
\usepackage{booktabs}
\usepackage{amsfonts}
\usepackage{multirow}
\usepackage{bbding}
\usepackage{cite}
\usepackage[urlcolor=blue]{hyperref}
\usepackage{subfigure}
\usepackage{cleveref}
\usepackage{siunitx}
\usepackage{mhchem}

\def\model{AV-SepFormer}

\title{AV-SepFormer: Cross-Attention SepFormer for Audio-Visual Target Speaker Extraction}
\name{
\begin{tabular}{@{}c@{}}Jiuxin Lin$^{1,\dagger}$\thanks{$^{\dagger}$ Work conducted when the first author was intern at Xiaomi Inc.}, Xinyu Cai$^{1}$, Heinrich Dinkel$^2$, Jun Chen$^1$, Zhiyong Yan$^2$, Yongqing Wang$^2$,\\
\textit{Junbo Zhang$^2$, Zhiyong Wu$^{1,3,*}$\thanks{$^{*}$ Corresponding author.}, Yujun Wang$^2$, Helen Meng$^{1,3}$}
\end{tabular}
}

\address{
    $^1$ 
    Shenzhen International Graduate School, Tsinghua University, Shenzhen, China\\
    $^2$ Xiaomi Inc., Beijing, China\\ 
    $^3$ 
         The Chinese University of Hong Kong, Hong Kong SAR, China\\
    \small{
        linjx21@mails.tsinghua.edu.cn, 
        dinkelheinrich@xiaomi.com,
        zywu@sz.tsinghua.edu.
    }
}
\begin{document}
\maketitle
\ninept
\begin{abstract}
Visual information can serve as an effective cue for target speaker extraction (TSE) and is vital to improving extraction performance. 
In this paper, we propose \model{}, a SepFormer-based attention dual-scale model that utilizes cross- and self-attention to fuse and model features from audio and visual.
\model{} splits the audio feature into a number of chunks, equivalent to the length of the visual feature.
Then self- and cross-attention are employed to model the multi-modal features.
Furthermore, we use a novel 2D positional encoding, that introduces the positional information between and within chunks and provides significant gains over the traditional positional encoding.
Our model has two key advantages:
the time granularity of audio chunked feature is synchronized to the visual feature, which alleviates the harm caused by the inconsistency of audio and video sampling rate; 
by combining self- and cross-attention, feature fusion and speech extraction processes are unified within an attention paradigm.
The experimental results show that \model{} significantly outperforms other existing methods.
\end{abstract}

\begin{keywords}
Target speech extraction, multi-modal fusion, cross-attention, 2D positional encoding.
\end{keywords}
\section{Introduction}
\label{sec:intro}

{T}{arget} speaker extraction (TSE) aims to extract a specific character's speech signal in the presence of multiple talkers and background noises. It has a wide range of real-world applications, such as hearing aids~\cite{wang2017deep}, speech recognition~\cite{zhang2018deep}, speaker verification~\cite{michelsanti2017conditional}, speaker diarization~\cite{sell2018diarization} and voice surveillance~\cite{geiger2015improving}.
Unlike blind speech separation (BSS), TSE methods are not restricted by permutation invariant training (PIT)~\cite{yu2017permutation} and require additional clue information to identify the target speaker.
Numerous techniques have been created by researchers to make use of various forms of cues, including reference speech~\cite{xu2020spex,ge2020spex+,ge2021multi}, still speaker face image~\cite{gao2021visualvoice}, and lip movement~\cite{wu2019time,pan2021muse,9887809,pan2022selective,gao2021visualvoice}.

Among these cues, lip movement is more advantageous than the others.
The reasons are as follows:
1) Audible background noise and reverberation have no impact on the visual lip movement information. 
2) Speech can be directly inferred from lip movement, and thus spoken content information can be effectively extracted.
3) Approaches that use visual cues are enrollment-free, making them easier to use in practice.
In this paper, we focus on enhancing the performance of lip movement based Audio-Visual TSE, for convenience we refer to this lip movement based Audio-Visual TSE as Audio-Visual TSE in the following.

Existing Audio-Visual TSE models typically up-sample the visual feature to align with the audio feature in time dimension~\cite{gao2021visualvoice, wu2019time}, which means these two modal features are fused without taking any difference in temporal granularity between visual and audio into account. This may lead to a degradation of TSE performance since the audio feature is fine-grained, whereas the visual feature tends to be coarse-grained.
Regarding the data types frequently employed in Audio-Visual TSE tasks, the face track video stream is available at 25 FPS (Frames Per Second), and the audio data commonly have a sampling rate of 8 or 16 kHz, which commonly is processed by a short time Fourier transform (STFT) to about 50 to 125 FPS~\cite{yang2022tfpsnet,wang2022tf}. Meanwhile, other time domain approaches using a trainable front-end instead of STFT commonly produce features at 200 FPS~\cite{pan2021muse,9887809}, leading to a more significant disparity of time granularity between audio and visual features.

Therefore, it is vital to consider integrating audio and visual features at the appropriate temporal granularity for Audio-Visual TSE models to perform even better.

To this end, we propose \model{}, a dual-scale model for TSE at both visual and audio scales. Inspired by DPRNN~\cite{luo2020dual} and SepFormer~\cite{subakan2021attention}, we split the audio feature sequence into shorter chunks such that the number of audio feature chunks matches the length of the visual feature sequence. In this way, the audio chunked feature has comparable temporal granularity to the visual feature. Three modules—IntraTransformer, InterTransformer and CrossModalTransformer, are interleaved for further modeling. The former two modules use self-attention at intra- and inter-chunk levels for short- and long-distance modeling, while the latter uses cross-attention for feature fusion at the inter-chunk level for the reason that long-term temporal context is essential to learn the audio-visual relationship~\cite{tao2021someone}.
Moreover, we introduce a 2D positional encoding method~\cite{raisi20202d} to help our attention-based model learn the 2D spatial relationship at intra- and inter-chunk levels, making it possible for cross-modal fusion to differentiate consecutive audio frames more clearly at the intra-chunk level. Experimental results on VoxCeleb2~\cite{chung2018voxceleb2}, LRS3~\cite{afouras2018lrs3} and TCD-TIMIT~\cite{harte2015tcd} datasets show that our \model{} consistently outperforms other existing advanced Audio-Visual TSE models in terms of signal and perceptual quality.

\section{Methodology}
\label{sec:format}
\begin{figure}
\centering
\includegraphics[width=0.6\linewidth]{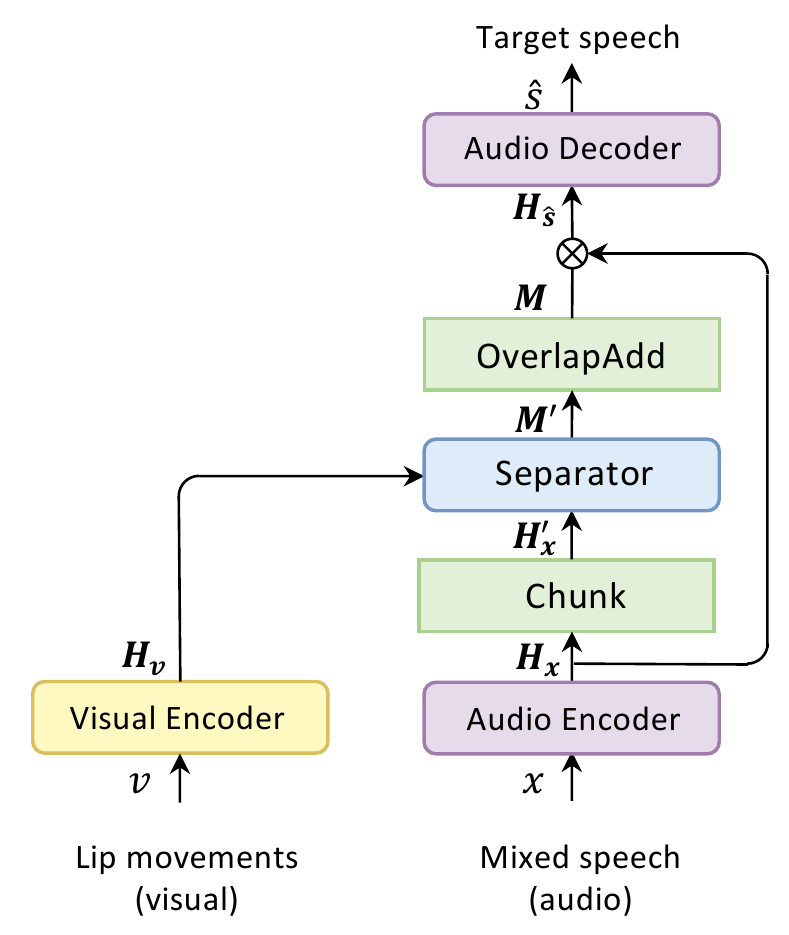}
\caption{ Overall structure of \model{}. \model{} consists of six parts: Visual Encoder, Audio Encoder, Chunk, Separator, OverlapAdd and Audio Decoder.}
\label{fig1}
\vspace{-1.5em}
\end{figure}
Following the paradigm of the time-domain TSE model~\cite{wu2019time}, our proposed \model{} employs an Audio Encoder, a Visual Encoder, an Audio Decoder, and a Separator. The overall structure of \model{} is shown in \Cref{fig1}.

\subsection{Audio Encoder}
\label{audio_encoder}
The Audio Encoder extracts the audio feature $ \boldsymbol{H_x}$ from a speech mixture $\boldsymbol{x} \in \mathbb{R}^{1 \times T}$ using the 1D convolution operation with a kernel size $L$ and stride $L/2$: 
\begin{equation}
 \boldsymbol{H_x}=\text{Conv1D}(\boldsymbol{x}, L, L/2)  \in \mathbb{R}^{N \times K},   
\end{equation}
where $N$ is the audio feature dimension and $K = \frac{T-L}{L / 2}+1$. This approach is also known as adaptive front-end~\cite{luo2020dual}, which seeks to replace the short-time Fourier transform (STFT) with a differentiable transform to learn a trainable feature representation from the .

For the 2D audio feature $\boldsymbol{H_x}$, Chunk operation splits $\boldsymbol{H_x}$ into chunks of length $C$ and hop size $C/2$. The padding operation is the same as which in SepFormer~\cite{subakan2021attention}. All chunks are then concatenated together to form a 3D audio chunked feature $\boldsymbol{H_x'}$:
\begin{equation}
 \boldsymbol{H_x'} = \textrm{Chunk}\left(\boldsymbol{H_x}\right) \in \mathbb{R}^{N \times C \times I},
\end{equation}
where $I$ indicates the number of chunks, it is designed to be exactly equivalent to the length of visual feature $\boldsymbol{H_v}$. 

\subsection{Visual Encoder}
\begin{figure}
\centering
\includegraphics[width=0.8\linewidth]{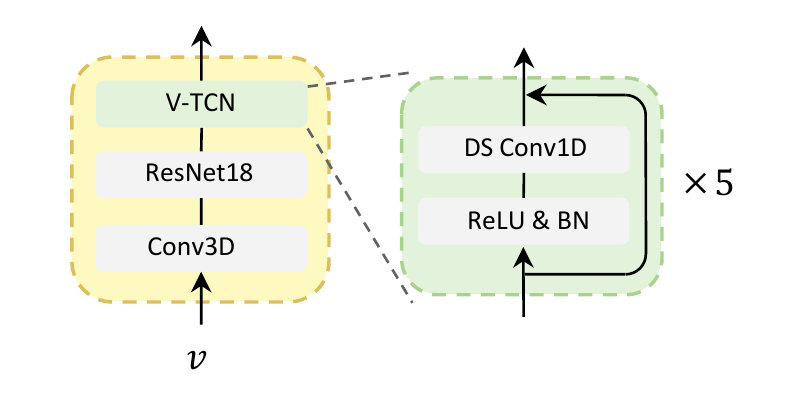}
\caption{Detailed structure of the Visual Encoder.}
\label{fig2}
\vspace{-1em}
\end{figure}

Our Visual Encoder follows previous Audio-Visual TSE~\cite{wu2019time,9887809,pan2021muse} methods, which use a pre-trained fixed-parameter lip embedding extractor consisting of a 3D convolution layer and an 18-layer ResNet~\cite{he2016deep} combined with multi-layer TCN networks.
Its detailed structure is shown in \Cref{fig2}.
Specifically, the Visual Encoder takes in the grey-scale image sequence $\boldsymbol{v}$ as input and outputs visual feature $\boldsymbol{H_v} \in \mathbb{R}^{N \times I}$,
where $N$ represents the visual feature dimension (equivalent to the audio feature dimension), and $I$ indicates the length of visual feature.
\subsection{Separator}
\begin{figure}
\centering
\includegraphics[width=1\linewidth]{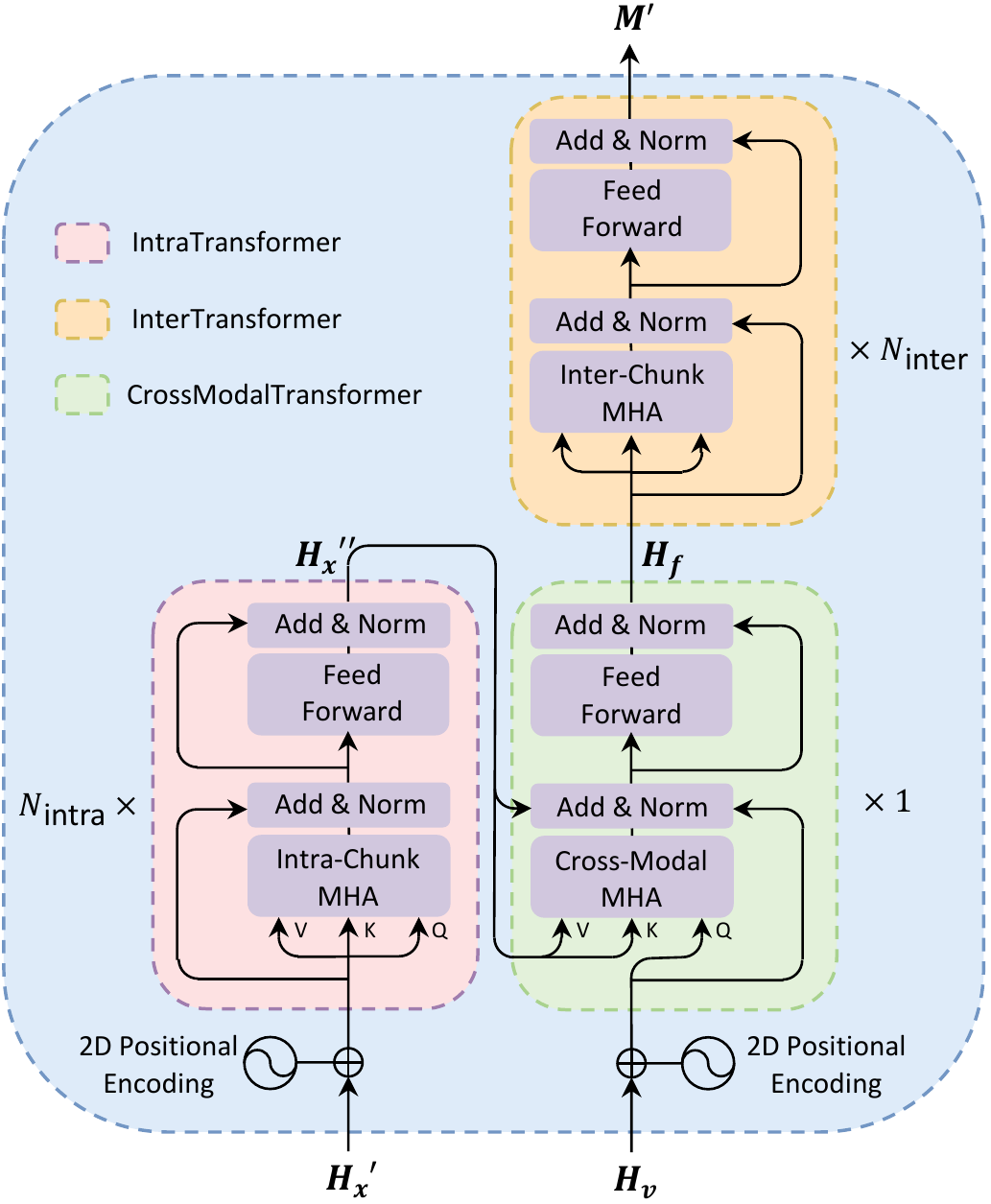}
\vspace{-1.5em}
\caption{Detailed structure of Separator. The whole Separator consists of three parts: IntraTransformer, InterTransformer and CrossModalTransformer.}
\label{fig3}
\vspace{-1.5em}
\end{figure}

As shown in \Cref{fig3}, the Separator of our \model{} consists of three parts: IntraTransformer, CrossModalTransformer and InterTransformer. The number of heads in each multi-head attention in these modules is abbreviated as $N_\text{head}$. IntraTransformer aims to model the audio chunked feature $\boldsymbol{H'_x}$ at fine-grained intra-chunk level. The audio chunked feature $\boldsymbol{H'_x}$ is progressively refined through $N_\text{intra}$ repeated blocks. Inside each block, self-attention is applied to each of the $C$ chunks respectively:
\begin{equation}
\boldsymbol{H''_x}[:,\ :,\ i] = {\textrm{IntraTransformer}}(\boldsymbol{H'_x}[:,\ :,\ i]), 
\end{equation}
where $\boldsymbol{H''_x}\in \mathbb{R}^{N \times C \times I}$ , $i \in \{1, 2, \cdots, I\}$.

CrossModalTransformer seeks to fuse audio and visual features at the same time granularity. Before the modal fusion phase, $\boldsymbol{H_v}$ is linearly converted to $\boldsymbol{Q_v}$, while $\boldsymbol{H_x''}$ is linearly converted to $\boldsymbol{K_x}, \boldsymbol{V_x}$:
\begin{equation}
\boldsymbol{Q_v} = \textrm{Linear}(\boldsymbol{H_v}) \in \mathbb{R}^{N \times I},
\end{equation}
\begin{equation}
\boldsymbol{K_{xc}},\ \boldsymbol{V_{xc}} = \textrm{Linear}(\boldsymbol{H_x''}[:,\ c,\ :]) \in \mathbb{R}^{N \times I},
\end{equation}
where $c \in \{1, 2, \cdots, C\}$. After that, $\boldsymbol{K_x}, \boldsymbol{V_x}, \boldsymbol{Q_x}$ are fed as the key, value and query respectively into CrossModalTransformer to output the fusion feature $\boldsymbol{H_f} \in \mathbb{R}^{N \times C\times I}$:
\begin{equation}
   \boldsymbol{H_f}[:,\ c,\ :] = \text{CrossModalTransformer}(\boldsymbol{Q}_v,\boldsymbol{K}_{xc},\boldsymbol{V}_{xc}) . 
\end{equation}

InterTransformer focuses on modeling the fusion feature at the coarse-grained inter-chunk level. Similar to the IntraTransformer, InterTransformer consists of $N_\text{inter}$ blocks, self-attention is applied inside each block. The difference is that we do self-attention for each slice on the $C$ dimension, and get chunked mask $\boldsymbol{M'}$:
\begin{equation}
\boldsymbol{M'}[:,\ c,\ :] = \textrm{InterTransformer}(\boldsymbol{H_f}[:,\ c,\ :]),
\end{equation}
where $\boldsymbol{M'} \in [0,1]^{N \times C \times I}$, $c \in\{1, 2, \cdots, C\}$.

\subsection{2D Positional Encoding}
\begin{figure}
\centering
\subfigure[1D positional encoding in \model{}]{\label{fig:1dpos}\includegraphics[width=1\linewidth]{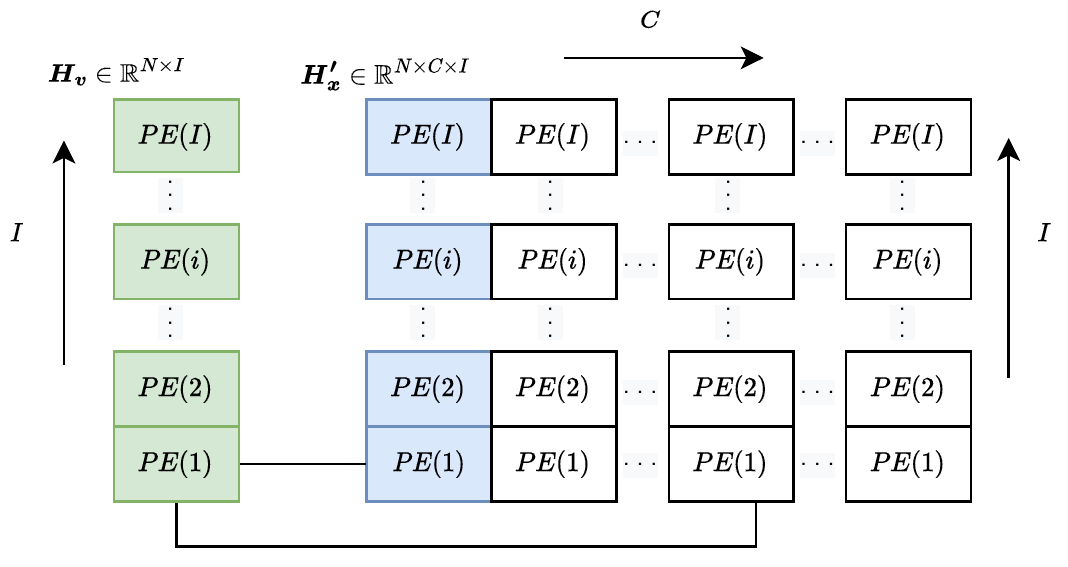}}
\vspace{-1.5em}
\subfigure[2D positional encoding in \model{}]{\label{fig:2dpos}\includegraphics[width=1\linewidth]{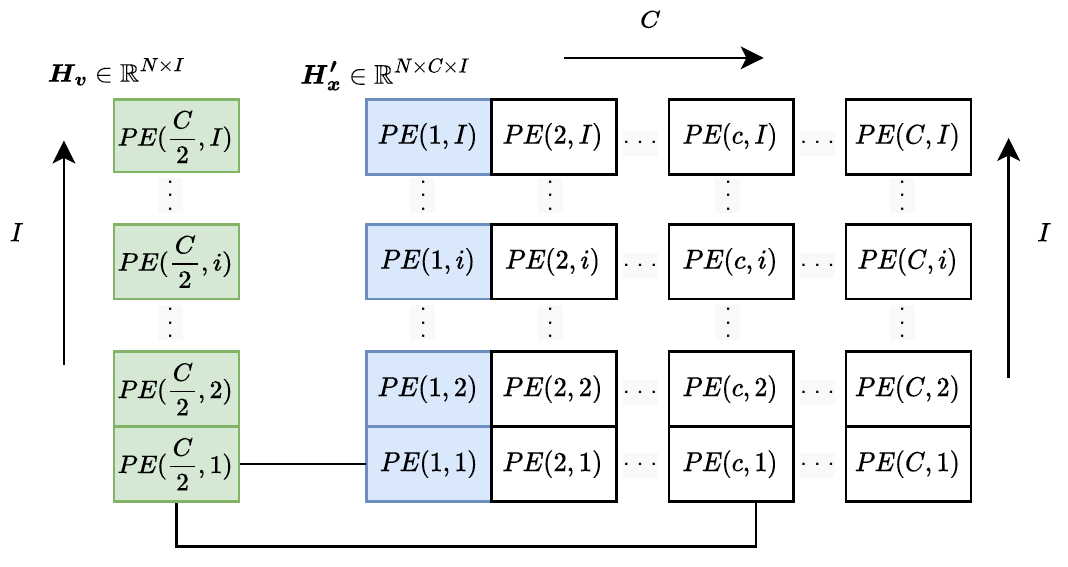}}
\caption{Difference between using 1D and 2D positional encoding in our \model{}. $PE( \cdot )$ means positional encoding. Every chunks' ($\boldsymbol{H_x'}$) positional encoding is identical in the 1D case. Note that this is a slice on feature dimension $N$.}
\label{fig:positional_embedding_explain}
\vspace{-1.5em}
\end{figure}

Positional encoding is considered essential when it comes to transformer-based models. 
However, the original positional encoding in~\cite{vaswani2017attention} is usually considered for encoding 2D features like $\boldsymbol{H_x}$ and $\boldsymbol{H_v}$. 
Due to the 3D shape of $\boldsymbol{H'_x}$, it is not applicable in \model{}, otherwise, positional encoding applied to every chunk of $\boldsymbol{H'_x}$ would be identical as is in \Cref{fig:1dpos}. This can not help us to take advantage of more detailed positional information inside every chunk, and thus may lead to a degradation of performance.


Thus we introduce 2D positional encoding~\cite{raisi20202d}, which uses the horizontal and vertical coordinates of the vectors in the 2D matrix to calculate the positional encoding values:
\begin{equation}
\begin{aligned}
& PE(c,\ i,\ 2u)=\sin \frac{c}{10000^{\frac{4u}{N}}} , \\
& PE(c,\ i,\ 2u+1)=\cos \frac{c}{10000^{\frac{4u}{N}}},\\
& PE\left(c,\ i,\ 2v+\frac{N}{2}\right)=\sin \frac{i}{10000^{\frac{4v}{N}}}, \\
& PE\left(c,\ i,\ 2v+1+\frac{N}{2}\right) =\cos \frac{i}{10000^{\frac{4v}{N}}},
\end{aligned}
\vspace{-0.05in}
\end{equation}
where $c \in\{1,2, \cdots, C\}$ and $i \in\{1,2, \cdots, I\}$ represent the position within and between the chunks, $u, \ v \in\left[0, \frac{N}{4}\right]$ denote the position in feature dimension.
For audio chunked feature $\boldsymbol{H'_x} \in \mathbb{R}^{N \times C \times I}$, simply add $PE(c, i, \cdot)$ to it by dimension. 
For visual feature $\boldsymbol{H_v} \in \mathbb{R}^{N \times I}$, due to its 2D shape, we choose to fix $c = \frac{C}{2}$ and remain $i$ corresponding to its $I$ dimension, add $PE(\frac{C}{2}, i, \cdot)$ to it by dimension at last. \Cref{fig:2dpos} shows the detailed process.
\vspace{-3mm}
\subsection{Audio Decoder}
OverlapAdd operation seeks to reconstruct 2D feature from chunked 3D feature. It can be regarded as the inverse of the Chunk operation in \Cref{audio_encoder}, 
\begin{equation}
  \boldsymbol{M} = \textrm{OverlapAdd}(\boldsymbol{M'})\in [0,1]^{N \times K}.
\end{equation}
The input to the Audio Decoder is the element-wise multiplication between the mask $\boldsymbol{M}$ of the target speaker and the output $\boldsymbol{H_x}$ of the Audio Encoder:
\begin{equation}
\vspace{-3mm}
\boldsymbol{{H}_{\hat{s}}}=\boldsymbol{H_x} \otimes \boldsymbol{M} \in \mathbb{R}^{N \times K},
\vspace{-3mm}
\end{equation}
where $\otimes$ means element-wise multiplication.
The Audio Decoder utilizes a transposed convolution layer, with the same stride and kernel size of the Audio Encoder in \Cref{audio_encoder} to reconstruct the target speech $\hat{s}$ from $\boldsymbol{H_{\hat{s}}}$:
\begin{equation}
\hat{s} = \textrm{TransposedConv1d}(\boldsymbol{{H}_{\hat{s}}}) \in \mathbb{R}^{1 \times T}.
\end{equation}
\vspace{-5mm}
\section{Experiments}
\vspace{-2mm}
\label{sec:pagestyle}

\subsection{Datasets}
We first evaluate \model{} on the VoxCeleb2 dataset~\cite{chung2018voxceleb2}.
The identities of speakers in the train and test sets have no overlap.
All utterances have a duration between 4 and 6 seconds. 
We select 48,000 utterances from 800 speakers in the train set and 36,237 utterances from 118 speakers in the test set. Then we simulate a 2-speaker mixture speech dataset of 20,000, 5,000, and 3,000 utterances for train, validation and test sets, respectively. Each interfering speech is mixed with the target speech at a random Signal-to-Noise ratio (SNR) set between -10 and 10 dB. We also compare \model{} with VisualVoice~\cite{gao2021visualvoice}, AV-ConvTasnet~\cite{wu2019time} and MuSE~\cite{pan2021muse} on cross-domain datasets, namely LRS3~\cite{afouras2018lrs3} and TCD-TIMIT~\cite{harte2015tcd}, which are created from TED videos and studio videos respectively. 

\begin{table}
\caption{Hyperparameters of \model{}}
\centering
\setlength{\tabcolsep}{3pt}
\resizebox{\linewidth}{!}{
\begin{tabular}{lccccc}
    \toprule
         Symbol & Description  & Value \\       
    \midrule
     $L$ & Kernel size in Conv1D of audio encoder  & 16  \\
    $C$   & Length of the chunk & 160  \\
     $N_\text{intra}$ & Number of the IntraTransformer & 8  \\
     $N_\text{inter}$ & Number of the InterTransformer  & 7  \\
      $N$ & Feature dimension of visual and audio feature  & 256  \\
      $N_\text{head}$ & Number of head in multi-head attention & 8 \\ 
    \bottomrule
\end{tabular}
}
\label{tab0}
\vspace{-2em}
\end{table}

\begin{table}
\caption{The comparison across various Audio-Visual TSE models on VoxCeleb2 2-speaker mixture.}
\setlength{\tabcolsep}{3pt}

\resizebox{\linewidth}{!}{
\begin{tabular}{lccccc}
    \toprule
         Network & Domain  & Auxiliary input & SI-SDR (dB) & PESQ \\       
    \midrule
     VisualVoice~\cite{gao2021visualvoice} & T-F & L(lip) + F(face)     & 9.732 & 1.965 \\
    AV-ConvTasNet~\cite{wu2019time} & T & L(lip) &10.381 & 1.973 \\
    AV-ConvTasnet (att) & T & L(lip)  &11.332 & 2.237 \\
    MuSE~\cite{pan2021muse} & T & L(lip) + S(label)   & 11.240  & 2.201 \\
    \model{} (ours) & T & L(lip)  & \textbf{12.130} & \textbf{2.313} \\
    \bottomrule
\end{tabular}
}
\label{tab1}
\vspace{-2em}
\end{table}
\subsection{Implementation setup}
We implement an \model{}\footnote{Code and demo available at \href{https://github.com/lin9x/AV-SepFormer}{https://github.com/lin9x/\model{}}} system as described in \Cref{sec:format}. The visual feature is extracted from the input video and resampled to 25 FPS. The audio is synchronized with the video and sampled at 16 kHz. The hyper-parameters of \model{} are shown in \Cref{tab0}. Note that here we fix $C=160$ to ensure that the audio chunked feature and the visual feature are  aligned. 

The model is trained using an Adam optimizer~\cite{kingma2015adam} with an initial learning rate of \num{1.5e-4}. The learning rate will get halved if there is no loss decrease on validation set for \num{3} epochs.
If there is no loss decrease for \num{5} epochs, the whole training process stops.
\vspace{-1em}
\subsection{Comparison with other state-of-the-art methods}
We first compare the objective performance of \model{} with other state-of-the-art methods on VoxCeleb2. The methods include one time-frequency domain method (VisualVoice~\cite{gao2021visualvoice}) and two time domain methods (AV-ConvTasnet~\cite{wu2019time}, MuSE~\cite{pan2021muse}). Note that time-frequency domain method means using STFT as the audio feature extractor to predict $\boldsymbol{M}$, while time-domain method replaces STFT with a learnable Audio Encoder as in~\ref{audio_encoder}. Scale-invariant signal-to-distortion ratio (SI-SDR) measures the speech signal quality, while perceptual evaluation of speech quality (PESQ)  measures the overall perceptual quality. They're all the higher, the better.

\Cref{tab1} presents the results of the comparison experiment, where ``T-F" means time-frequency method, while ``T" stands for time domain method. In addition to the use of lip information as an auxiliary input, VisualVoice~\cite{gao2021visualvoice} also uses extra information about the speaker's face as an auxiliary input. Meanwhile, MuSE~\cite{pan2021muse} requires an additional speaker label, while most models require only visual features from a pre-trained extractor. Results demonstrate that our proposed \model{} has a relative improvement of 0.9 dB and 0.1 in terms of SI-SDR and PESQ respectively, over MuSE. Besides, we employ cross-attention in AV-ConvTasNet (att) to fuse the audio and visual features~\cite{sato2021multimodal} instead of direct concatenation in the original AV-ConvTasnet. It's worth pointing out that \model{} outperforms the AV-ConvTasnet (att) by 0.8 dB and 0.1 in terms of SI-SDR and PESQ. This is due to the dual-scale design of \model{} helping to solve the problem of different temporal granularity between the audio and visual features and the Transformer's superior sequence modeling capability over TCN.

On LRS3 and TCD-TIMIT, \model{} still shows strong competitiveness compared to other models, and results are shown in \Cref{tab2}.

\begin{table}
\centering
\caption{Cross-datasets evaluations of various Audio-Visual TSE models. Models are trained on VoxCeleb2 and tested on LRS3 and TCD-TIMIT.}

\resizebox{\linewidth}{!}{
\begin{tabular}{lcccc}
\toprule
 \multirow{2}{1cm}{Network}  & \multicolumn{2}{c}{ LRS3 } & \multicolumn{2}{c}{ TCD-TIMIT } \\
  \cmidrule(lr){2-3} \cmidrule(lr){4-5}
& SI-SDR (dB) & PESQ & SI-SDR (dB) & PESQ \\
   \midrule 
 VisualVoice~\cite{gao2021visualvoice} & 11.603 & 2.269 & 10.883 & 2.246 \\
AV-ConvTasNet~\cite{wu2019time} & 12.128 & 2.326 & 11.532 & 2.210 \\
MuSE~\cite{pan2021muse} & 12.971 & 2.558 & 12.497 & 2.447 \\
\model{} (ours)  & \textbf{13.815} & \textbf{2.668} & \textbf{13.436} & \textbf{2.567} \\
\bottomrule
\end{tabular}
}
\label{tab2}
\vspace{-2em}
\end{table}

\begin{table}
\centering
\caption{The ablation study on VoxCeleb2 dataset, CA and 2DPos means cross-modal attention and 2D positional encoding respectively.}
\resizebox{\linewidth}{!}{
\begin{tabular}{lccccc}
\toprule
 Models & CA &  2DPos& SI-SDR (dB) & PESQ  \\
  \midrule 
\model{} & \CheckmarkBold & \CheckmarkBold & \textbf{12.130} & \textbf{2.313}  \\
$\text{-}\ \text{w/o}$ CA       &   \XSolidBrush  & \CheckmarkBold  & 11.701 & 2.268 \\
$\text{-}\ \text{w/o}$ 2DPos    &  \CheckmarkBold &  \XSolidBrush  & 12.020 & 2.295 \\
$\text{-}\ \text{w/o}$ CA \& 2DPos    & \XSolidBrush & \XSolidBrush & 11.522 & 2.261\\
\bottomrule
\end{tabular}
}
\label{tab3}

\end{table}
\vspace{-1.5em}
\subsection{Ablation study}
\label{ablation}
To determine the effectiveness of the improved method proposed in this paper, we study variants of \model{}. \Cref{tab3} shows the performances of these variants, where ``w/o" means without, ``CA" and ``2DPos" are the abbreviations of cross-modal attention and 2D positional encoding, respectively. 

Results show that after using cross-modal attention instead of directly concatenating different modal features, the model can obtain about 0.4 dB and 0.05 performance improvement in SI-SDR and PESQ. Using 2D instead of 1D positional encoding improves the model performance by about 0.1 dB and 0.02 in terms of SI-SDR and PESQ. Using both methods, the model achieves about 0.61 dB and 0.05 in terms of SI-SDR and PESQ compared to the baseline model. The above experiments demonstrate that cross-modal attention (CA) and 2D positional encoding (2DPos), alone and in combination, can lead to a significant improvement in the performance of \model{}.

\vspace{-3mm}
\section{Conclusions}
\vspace{-2mm}
\label{sec:conclusions}
In this paper, we propose a dual-scale transformer-based model \model{} to overcome the degradation of Audio-Visual TSE tasks caused by temporal granularity misalignment of audio and visual features. Specifically, we employ CrossModalTransformer to fuse the multi-modal features at coarse-grained level. Fine- and coarse-grained feature information is modeled by IntraTransformer and InterTransformer, respectively. Furthermore, we introduce 2D positional encoding to help our model focus more on the positional relationships of the audio chunked feature at intra- and inter-chunk levels. Experimental results demonstrate the effectiveness of the proposed \model{} with a combination of the above three modules and the special positional encoding. We also compare \model{} with other top-ranked Audio-Visual TSE methods on three public datasets, which shows the superior performance of the proposed \model{} on both speech signal quality and overall perceptual quality.

\textbf{Acknowledgements:}
This work is supported by National Natural Science Foundation of China (62076144), Major Key Project of PCL (PCL2021A06, PCL2022D01), AMiner.Shenzhen SciBrain.

\vfill\pagebreak

\ninept
\bibliographystyle{IEEEbib}
\bibliography{main}

\end{document}